\documentstyle[aps,twocolumn,tighten,epsfig]{revtex}
\begin{document}

\draft
\title{ DENSITY OF HEDGEHOGS JUST AFTER A QUENCH }

\author{Jacek Dziarmaga\thanks{E-mail: {\tt ufjacekd@thrisc.if.uj.edu.pl}}}
\address{ Institute of Physics, Jagiellonian University,
          Reymonta 4, 30-059 Krak\'ow, Poland}
\date{June 11, 1998}
\maketitle
\tighten

\begin{abstract}
{\bf Density of vortices and monopoles after a rapid pressure
quench is predicted. Critical exponents are worked out in overdamped
and underdamped limits. }
\end{abstract}
\vspace*{0.5cm}

\section{ Introduction. }

  Topological defects play a prominent role in many condensed matter systems,
see e.g.\cite{davydov} for a review. It was also suggested that they
were an important ingredient of the early universe \cite{vilenkin}.
Topological defects can be generated in large numbers during
a second order phase transition. The dynamics of such a transition
to a symmetry broken phase has been an object of much recent attention
because of its importance in the cosmological context \cite{cosmos} and in
condensed matter physics \cite{condmat}. An early estimate of defect density
after a quench was given by Kibble \cite{kibble}. It this theory, the speed
of light is a dominant factor which determines the size of correlated domains.
A more detailed scenario was put forward by Zurek \cite{zurek}, who
emphasized the importance of the nonequilibrium dynamics of the order
parameter. Recent experiments point to the latter theory.

  In a recent paper \cite{jd2} we calculated density of kinks in an 
underdamped one dimensional $\phi^4$ theory.
In this paper we give a comprehensive treatment of hedgehogs in
$D=2,3$ spatial dimensions. A hedgehog is a point like defect which
is topologically stable thanks to its asymptote at infinity.
The asymptote is a topologically nontrivial map $S^D \rightarrow S^D$.
A hedgehog should be distinguished from a texture the asymptote of which
is a constant. The paper has essentially two parts. In the first we give 
a general outline of the formalism. In the second part the formalism is 
applied to vortices and monopoles.

\section{ Model. }

  Let us consider an $O(D)$ symmetric time dependent Ginzburg-Landau model
in $D$ spatial dimensions

\begin{eqnarray}\label{model}
&& \ddot{\phi^{\alpha}}(t,\vec{x})+
   \gamma\dot{\phi^{\alpha}}(t,\vec{x}) -
   \frac{\partial}{\partial x^k}\frac{\partial}{\partial x^k}
   \phi^{\alpha}(t,\vec{x}) =               \nonumber \\
&& - 2 a(t) \phi^{\alpha}(t,\vec{x})
   - 2[\phi^{\beta}(t,\vec{x})\phi^{\beta}(t,\vec{x})]
   \phi^{\alpha}(t,\vec{x})+
   \eta^{\alpha}(t,\vec{x})    \;\;,
\end{eqnarray}
where $k=1,\ldots,D$ are spatial indices and $\alpha,\beta=1,\ldots,D$
are internal indices. Repeated indices imply summation.
$\phi^{\alpha}(t,\vec{x})$'s are real. $\eta^{\alpha}(t,x)$'s are Gaussian
white noises of temperature $T$ with correlations

\begin{eqnarray}\label{correlations}
&& <\eta^{\alpha}(t,\vec{x})>=0 \;\;, \nonumber\\
&& <\eta^{\alpha_1}(t_1,\vec{x}_1)\eta^{\alpha_2}(t_2,\vec{x}_2)>=\nonumber\\
&&\;\;\;\;2\gamma T\; \delta_{\alpha_1 \alpha_2}
                    \; \delta(t_1-t_2)
                    \; \delta^{(D)}(\vec{x}_1-\vec{x}_2) \;\;.
\end{eqnarray}
The coefficient $a(t)$ is time dependent. We consider a linear quench

\begin{equation}\label{a(t)}
a(t)=
\left\{
\begin{array}{ll}
A                   & \mbox{ , if $\; t<0$ }                   \\
A(1-\frac{t}{\tau}) & \mbox{ , if $\; 0<t<\tau\frac{A+1}{A}$ } \\
-1                  & \mbox{ , if $\; \tau\frac{A+1}{A}<t$ }
\end{array}
\right. 
\end{equation}
Before the quench, for $t<0$, the system is in a symmetric phase $(A>0)$,
during the quench, at $t=\tau$, it undergoes a transition from the symmetric
phase $a(t<\tau)>0$ to a broken symmetry phase $a(t>\tau)<0$. Finally it
settles down at $a(t)=-1$.

\section{ Soliton Transform. }

  The model (\ref{model}) in its symmetry broken phase ($a(t)=-1$) admits
pointlike topological defects or solitons

\begin{equation}\label{soliton}
\phi^{\alpha}(\vec{x})=S^{\alpha}(\vec{x}-\vec{z}).
\end{equation}
$\vec{z}$ is an arbitrary location of the soliton. The soliton solution
has $D$ zero modes

\begin{equation}\label{zeromodes}
Z^{\alpha}_{k}(\vec{x}-\vec{z})=
-\frac{\partial}{\partial z^k} S^{\alpha}(\vec{x}-\vec{z})=
 \frac{\partial}{\partial x^k} S^{\alpha}(\vec{x}-\vec{z})  \;\;.
\end{equation}
When field equations are linearized around the soliton solution
(\ref{soliton}), one obtains a discrete/continous spectrum
of eigenmodes of the fluctuation operator. There are $D$ zero modes
(\ref{zeromodes}) and a spectrum of modes with positive eigenvalues
and eigenfunctions $P^{\alpha}_{\rho}(\vec{x}-\vec{z})$ labelled
by a discrete/continuous index $\rho$. A field
$\phi^{\alpha}(t,\vec{x})$ can be expanded as a sum of
the soliton solution (\ref{soliton}) with an appropriate
choice of the soliton position $\vec{z}$ and a combination of
positive eigenvalue eigenmodes with coefficients $\Phi_{\rho}$,

\begin{equation}\label{expansion}
\phi^{\alpha}(t,\vec{x})=S^{\alpha}[\vec{x}-\vec{z}(t)]+
\sum_{\rho\neq 0} \Phi_{\rho}(t)
                  P^{\alpha}_{\rho}[\vec{x}-\vec{z}(t)] \;\;.
\end{equation}
$\rho\neq 0$ in the sum stresses that the zero modes are excluded
from the expansion, they are absorbed into the soliton location
$\vec{z}(t)$. Given the order parameter $\phi^{\alpha}(t,\vec{x})$,
one can trace the soliton position $\vec{z}(t)$ with a help of
a "soliton transform" of the order parameter

\begin{equation}\label{solitontranform}
I^{k}(t,\vec{x})=
\int d^{D}y\; Z^{\alpha}_{k}(\vec{y}) \;
              \phi^{\alpha}(t,\vec{x}+\vec{y}) \;\;.
\end{equation}
$I^{k}[t,\vec{z(t)}]=0$ for the field (\ref{expansion}) because
the zero modes (\ref{zeromodes}) are orthogonal to the soliton
solution (\ref{soliton}) and to the eigenmodes
$P^{\alpha}_{\rho}(\vec{x}-\vec{z})$.

  The soliton position $\vec{z}(t)$ is a zero of the soliton
tranform, $I^{k}[t,\vec{z(t)}]=0$, \cite{jw}.

  It may happen that there are several solitons in the system.
In that case an expansion like (\ref{expansion}) is possible around
any soliton position $\vec{z}_{1}(t),\vec{z}_{2}(t),\ldots$. Each
$\vec{z}(t)$ is a zero of the soliton transform. To count the solitons
one should count the zeros of the soliton transform. An average density
of solitons can be found as

\begin{equation}
n(t)
 =\int_{L^D} \frac{d^{D}x}{L^D}\;
  < \delta^{(D)}[ \vec{I}(t,\vec{x}) ]\;
    |\;\frac{\partial (I^{1},\ldots,I^{D}) }
         {\partial (x^{1},\ldots,x^{D}) }(t,\vec{x})\;|\; > \;\;,
\end{equation}
where the integration goes over a $D$-dimentional box of size $L$.
Following similar steps as in \cite{halperin}, we obtain

\begin{eqnarray}\label{density}
&& n(t) = C_D
          (\frac{-W''(t,r=0)}{ W(t,r=0) })^{\frac{D}{2}}\;\;,
                                                      \nonumber\\
&& W(t,|\vec{x}|)=<I^k(t,0) I^k(t,\vec{x})> \;\;,     \nonumber\\
&& C_D =
   \left\{
   \begin{array}{ll}
   \frac{1}{\pi}       & \mbox{ , if $\; D=1 $ } \\
   \frac{1}{2\pi}      & \mbox{ , if $\; D=2 $ } \\
   \frac{1}{\pi^2}     & \mbox{ , if $\; D=3 $ }
   \end{array}
   \right.
\end{eqnarray}
Once the correlation function $W(t,r)$ is known one can work out
the average density of solitons in the system. In the considered
cases the soliton transform does not distinguish solitons from
antisolitons so Eq.(\ref{density}) gives a total density of
topological defects.

  It is convenient to express $W(t,r)$ by a correlation function
of the order parameter. The order parameter can be expanded in
Fourier modes

\begin{equation}\label{fourierexpansion}
\phi^{\alpha}(t,\vec{x})=
\int d^D p\; e^{i\vec{p}\vec{x}} \tilde{\phi}^{\alpha}(t,\vec{p}) \;\;.
\end{equation}
Thanks to translational invariance and the $O(D)$ symmetry of the
model (\ref{model}), a correlation function of the Fourier modes has
the form

\begin{equation}\label{Gdefinition}
<\tilde{\phi}^{\alpha_1}(t,\vec{p_1})
 \tilde{\phi}^{\alpha_2}(t,\vec{p_2})>\equiv
 \frac{ T \delta^{\alpha_1 \alpha_2}
          \delta^{(D)}(\vec{p}_1-\vec{p}_2)
          G(t,p_1)                             }
      {  (2\pi)^D                              }.
\end{equation}
This definition and the Fourier transform of the zero mode,

\begin{eqnarray}
&& \tilde{Z}^{\alpha}_{k}(\vec{p})=
   \int d^{D}x\; e^{-i\vec{p}\vec{x}} Z^{\alpha}_{k}(\vec{x}) \;\;,\nonumber\\
&& U(|\vec{p}|)=\tilde{Z}^{\alpha}_{k}(+\vec{p})
                \tilde{Z}^{\alpha}_{k}(-\vec{p}) \;\;,
\end{eqnarray}
leads to the following expression for the density of solitons

\begin{equation}\label{n}
n(t) =
  C_D
  (\frac{ \int d^D p\; p^2\;U(p)\;G(t,p) }
        { \int d^D p\; U(p)\;G(t,p) } )^{\frac{D}{2}}\;\;.
\end{equation}

  To find the density of solitons just after the quench (\ref{a(t)})
we need a correlation function $G(\tau,p)$ when the parameter
$a(t)$ passes through the critical point and enters the symmetry broken
phase. This correlation function is a topic of the next section.

\section{ Correlation Function At The Critical Point. }

  As long as the system is still in the symmetric phase ($t<\tau$),
the order parameter $\phi^{\alpha}$ can be regarded as a small fluctuation
around its symmetric ground state $<\phi^{\alpha}(t,\vec{x})>=0$. It is
justified to neglect the cubic term on the RHS of Eq.(\ref{model}).
A Fourier transform of the linearized Eq.(\ref{model}) is

\begin{eqnarray}\label{modelfourier}
&&\ddot{\tilde{\phi}}^{\alpha}(t,\vec{p}) +
  \gamma\dot{\tilde{\phi}}^{\alpha}(t,\vec{p})=
  -[ 2 a(t) + p^2 ]\tilde{\phi}^{\alpha}(t,\vec{p})+
  \tilde{\eta}^{\alpha}(t,\vec{p})       \;\;,\\
&& <\tilde{\eta}^{\alpha}(t,\vec{p})>=0  \;\;,\nonumber \\
&& <\tilde{\eta}^{\alpha_1}(t_1,\vec{p}_1)
    \tilde{\eta}^{\alpha_2}(t_2,\vec{p}_2)>=  \nonumber \\
&& \;\;\;\;
   \frac{2\gamma T}{ (2\pi)^D }\;
   \delta^{\alpha_1 \alpha_2}\;
   \delta(t_1-t_2)\;
   \delta^{(D)}(\vec{p}_1-\vec{p}_2) \;\;.    \nonumber \\
\end{eqnarray}

The correlation function $G(\tau,p)$ can be easily worked
out in two opposite limits.

\subsection{Underdamped limit, $\gamma\rightarrow 0$.}

   The underdamped limit is achieved if $2a(t)>>\gamma^{2}/4$
from the beginning of the quench to the freeze-in time $\hat{t}$.
In this limit the system (\ref{modelfourier}) can be mapped to the
equivalent model

\begin{eqnarray}\label{model2}
&& \dot{\tilde{\phi}}^{\alpha}(t,\vec{p})=
   -\frac{\gamma}{2}\tilde{\phi}^{\alpha}(t,\vec{p})+
   n^{\alpha}(t,\vec{p}) \;\;, \nonumber \\
&& <n^{\alpha}(t,\vec{p})>=0  \;\;,\nonumber \\
&& <n^{\alpha_1}(t_1,\vec{p}_1) n^{\alpha_2}(t_2,\vec{p}_2)>= \nonumber\\
&&\;\;\;\;   \frac{2\gamma T\;\delta^{\alpha_1 \alpha_2}\;
             \delta(t_1-t_2)\; \delta^{(D)}(\vec{p}_1-\vec{p}_2) }
             {(2\pi)^D [2a(t_1)+p^2_1]}\;\;.
\end{eqnarray}
To see the nature of this equivalence let us consider three examples.
If $a(t)=const$, then both systems have the same correlation

\begin{equation}\label{Gequilibrium}
G(t,p)=\frac{1}{2a+p^2}      \;\;.
\end{equation}
Their response to the change of parameters is also the same. In particular
they have the same relaxation time. For example, if $a(t)=const$ and the
temperature jumps from $T=0$ for $t<0$ to $T>0$ for $t>0$, then the
correlation function for $t>0$ is

\begin{eqnarray}
&&G(t,p)=\frac{1}{2a+p^2} [1-e^{-\gamma t}] \;\;,\nonumber\\
&&G(t,p)=\frac{1}{2a+p^2}
[1-e^{-\gamma t}+e^{-\gamma t}O(\frac{\gamma}{\sqrt{2a+p^2}})]
\end{eqnarray}
for the models (\ref{model2}) and (\ref{modelfourier}) respectively.
$O(\ldots)$ are oscillating terms which
are negligible for small $\gamma$. In this limit the response
of the two systems is the same. Another example is a response
to a sudden change in the parameter $a$ from $a=a_{-}$
for $t<0$ to $a=a_{+}$ for $t>0$

\begin{eqnarray}
G(t,p)&=&\frac{1-e^{-\gamma t}}{2a_{+}+p^2}+
          \frac{e^{-\gamma t}}{2a_{-}+p^2}    \;\;,\nonumber\\
G(t,p)&=&\frac{1-e^{-\gamma t}}{2a_{+}+p^2}+
            \frac{e^{-\gamma t}}{2a_{-}+p^2}   +   \nonumber\\
      & & O(\frac{\gamma e^{-\gamma\tau}}{\sqrt{2a_{+}+p^2}},
            \frac{\gamma e^{-\gamma\tau}}{\sqrt{2a_{-}+p^2}})    \;\;,
\end{eqnarray}
where $O(\ldots)$ are once again oscillations negligible
for $\gamma\rightarrow 0$. The two considered systems respond in
the same way to the impulsive perturbations of their parameters. The
impulsive perturbations contain all frequencies so the models
can be considered equivalent in the underdamped limit.

   In the underdamped limit the relaxation times of all the Fourier modes
are the same and equal to $1/\gamma$. One can talk about a relaxation time
of the system as a whole. As $a(t)$ begins to change at $t=0$ the system
tries to adapt to the changes. It can not react faster than on the time
scale of its relaxation time $1/\gamma$. The correlations are effectively
frozen at the freeze-in time $\hat{t}$, when the time left to the transition
is equal to the relaxation time, $(\tau-\hat{t})=\frac{1}{\gamma}$.
At this instant $a(\hat{t})=\frac{A}{\gamma\tau}$, so the frozen
correlation function is

\begin{equation}\label{Gfrozenunder}
\hat{G}_{\gamma\rightarrow 0}(\tau,p)=\frac{1}{\frac{2A}{\gamma\tau}+p^2}      \;\;.
\end{equation}
The frozen correlation length can be identified as

\begin{equation}\label{xifrozenunder}
\hat{\xi}_{\gamma\rightarrow 0}=
  \sqrt{\frac{\gamma\tau}{2A}}\sim
  \tau^{1/2} \;\;.
\end{equation}
These asymptotic results are valid for $\tau>>1/\gamma$. The system
remains in the underdamped limit until the freeze-in time,
$2a(t<\hat{t})>>\gamma^2/4$, if  the condition

\begin{equation}
8 A > \tau \gamma^3
\end{equation}
is satisfied. If $\tau<<1/\gamma$, then the system is frozen from the
beginning of the quench and the frozen correlation length is
$\hat{\xi}=1/\sqrt{2A}$.

   The analysis so far was on a heuristic level. The equation (\ref{model2})
is of first order in time, so the exact correlation function of
the system (\ref{model2}) can be easily worked out. The formal solution of
the model (\ref{model2}) is

\begin{equation}
\tilde{\phi}^{\alpha}(\tau,\vec{p})=
\int_{-\infty}^{\tau}dt_1\;
e^{-\gamma (\tau-t_1)/2} n^{\alpha}(t_1,p).
\end{equation}
With this solution the correlation function is

\begin{equation}\label{Gexactunder}
G(\tau,p)=\frac{e^{-\gamma\tau}}{2A+p^2}+
\int_{0}^{\gamma\tau}dx\; \frac{e^{-x}}{\frac{2A}{\gamma\tau}x+p^2} \;\;.
\end{equation}
It turns out that Eq.(\ref{Gfrozenunder}) is indeed the asymptote of
the exact solution for $\tau>>1/\gamma$.

\subsection{ Overdamped limit, $\gamma\rightarrow\infty$. }

  In the overdamped limit ($\gamma^2/4>>2a(t)$) the second order time
derivative on the LHS of Eqs.(\ref{model},\ref{modelfourier}) can be
neglected. Eq.(\ref{modelfourier}) is simplified to

\begin{equation}\label{modelover}
\gamma\dot{\tilde{\phi}}^{\alpha}(t,\vec{p})=
   -[ 2 a(t) + p^2 ]\tilde{\phi}^{\alpha}(t,\vec{p})+
   \tilde{\eta}^{\alpha}(t,\vec{p})\;\;.
\end{equation}
In an overdamped system a relaxation time of a given Fourier
mode depends on its momentum $p$ and equals $\gamma/2[2a(t)+p^2]$. There is
no universal relaxation time of the system as a whole. Short wave lengths
freeze in later than the long wave lengths. For a given momentum $p$,
the relaxation time is equal to the time left to the transition,
if $(\tau-\hat{t}_p)=\gamma/2[2a(\hat{t}_p)+p^2]$. On this heuristic
level the long wave length limit of the frozen correlation function is

\begin{equation}
\hat{G}_{heuristic}(\tau,p)\approx
\frac{1}{\sqrt{\frac{\gamma A}{\tau}}+p^2} \;\;.
\end{equation}
The frozen correlation length can be identified as

\begin{equation}
\hat{\xi}_{heuristic}=
(\frac{\tau}{\gamma A})^{\frac{1}{4}}\sim
\tau^{1/4} \;\;.
\end{equation}

  The model (\ref{modelover}) has an exact solution

\begin{equation}
\tilde{\phi}^{\alpha}(\tau,\vec{p})=
\int_{-\infty}^{\tau}dt_1\;
  \exp\{ -\int_{t_1}^{\tau} dt_2 \; [2 a(t_2) + p^2]/\gamma \}\;
  \tilde{\eta}(t_1,\vec{p}) \;\;.
\end{equation}
This solution, the correlations in Eq.(\ref{modelfourier}) and the
explicit form of $a(t)$, see Eq.(\ref{a(t)}), give an exact correlation
function at $t=\tau$

\begin{eqnarray}\label{Gexactover}
&& G(\tau,p) = \frac{e^{-2 p^2 \tau/\gamma}}{2A+p^2} + \nonumber\\
&& \sqrt{\frac{\pi\tau}{2\gamma A}}
   e^{\frac{\tau p^4}{2\gamma A}}
  [Erf(\sqrt{\frac{\tau}{2\gamma A}}\;(2A+p^2))
   -Erf(\sqrt{\frac{\tau}{2\gamma A}}\;(p^2))] \;\;.
\end{eqnarray}
If $\tau$ is much longer than the relaxation time of the $p=0$
mode before the quench, $\tau>>2\gamma A$, the correlation function
takes an asymptotic form

\begin{equation}\label{Gasymptover}
G(\tau,p)\approx
  \sqrt{\frac{2\tau}{\gamma A}}
  e^{ \frac{p^4 \tau}{2\gamma A} }
  \int_{\sqrt{\frac{\tau}{2\gamma A}}p^2}^{+\infty} dx\;e^{-x^2} \;\;.
\end{equation}
The frozen correlation length can be identified as

\begin{equation}\label{xifrozenover}
\hat{\xi}_{\gamma\rightarrow\infty}=
(\frac{\tau}{\pi\gamma A})^{\frac{1}{4}} \;\;.
\end{equation}

\section{ Density of vortices. }

  In two dimensions ($D=2$) the model (\ref{model}) admits a topologically
stable vortex solution

\begin{equation}
S^{\alpha}(\vec{x})=f(\rho) [\delta^{\alpha 1}\cos(\varphi)
                            +\delta^{\alpha 2}\sin(\varphi)] \;\;,
\end{equation}
where $\rho,\varphi$ are polar coordinates and the profile function
interpolates between $f(0)=0$ and $f(\infty)=1$. A single vortex
has infinite energy, the IR divergence is due to the long distance
asymptote

\begin{equation}\label{asymptvortex}
S^{\alpha}_{\infty}(\vec{x})=[\delta^{\alpha1}\cos(\varphi)
                             +\delta^{\alpha2}\sin(\varphi)] \;\;.
\end{equation}
The asymptote is scale invariant. The asymptote determines the
long wave length asymptote of the Fourier transform, which is

\begin{equation}\label{Uvortex}
U(p)\sim \frac{1}{p^2} \;\;
\end{equation}
up to a numerical factor.

\subsection{Underdamped limit.}

  Combination of the formulas (\ref{Gfrozenunder},\ref{n},\ref{Uvortex})
should give the density of vortices for large $\tau$. The numerator
integral in (\ref{n}) is proportional to

\begin{eqnarray}
&& \int_{\Lambda}^{+\infty} p\;dp\;p^2\; U(p)\; G(\tau,p)
   \stackrel{\tau\rightarrow\infty}{\approx}                     \nonumber\\
&& \int_{\Lambda}^{+\infty} p\;dp\;p^2\; U(p)\;
   \frac{1}{\frac{2A}{\gamma\tau}+p^2}
   \stackrel{\tau\rightarrow\infty}{\approx}                     \nonumber\\
&& \int_{\Lambda}^{\infty} p\;dp\; U(p)\;
   \stackrel{\Lambda\rightarrow 0}{\approx}-\ln(\Lambda)         \;\;.
\end{eqnarray}
$\Lambda$ is an IR regulator. The denominator integral is proportional to

\begin{eqnarray}
&& \int_{\Lambda}^{+\infty} p\;dp\; U(p)\; G(\tau,p)
   \stackrel{\tau\rightarrow\infty}{\approx} \nonumber\\
&& \int_{\Lambda}^{+\infty} p\;dp\; U(p)\;
   \frac{1}{\frac{2A}{\gamma\tau}+p^2}
   \stackrel{\Lambda\rightarrow 0}{\approx}                      \nonumber\\
&& \frac{\gamma\tau}{2A} \int_{\Lambda} \frac{dp}{p}=
   -\frac{\gamma\tau}{2A} \ln(\Lambda)                           \;\;.
\end{eqnarray}
We manipulated with the order of the limits $\tau\rightarrow\infty$
and $\Lambda\rightarrow 0$ so as to get rid of the IR divergence
and obtain a finite regularized asymptote

\begin{equation}
n_{REG.} \approx \frac{A}{\pi\gamma\tau}  \;\;.
\end{equation}
The results for vortices are dominated by the IR divergence, which
is due to the asymptote (\ref{asymptvortex}). The asymptote is scale
invariant so the density of vortices is just an inverse squared of the 
frozen correlation length (\ref{xifrozenunder}).

\subsection{Overdamped limit.}

   In this limit we combine the formulas
(\ref{Gasymptover},\ref{n},\ref{Uvortex}). The numerator in Eq.(\ref{n})
is proportional to

\begin{eqnarray}
&& \int_{\Lambda}^{+\infty} p\;dp\;p^2\; U(p)\; G(\tau,p)
   \stackrel{\tau\rightarrow\infty}{\approx}                \nonumber\\
&& \int_{\Lambda}^{+\infty} p\;dp\; U(p)\;
   \stackrel{\Lambda\rightarrow 0}{\approx}
   -\ln(\Lambda) \;\;.
\end{eqnarray}
The denominator is

\begin{eqnarray}
&& \int_{\Lambda}^{+\infty} p\;dp\; U(p)\; G(\tau,p)
   \stackrel{\tau\rightarrow\infty}{\approx}                \nonumber\\
&& \sqrt{\frac{2\tau}{\gamma A}}
   \int_{\Lambda}^{+\infty} p\;dp\;U(p)\;
                            e^{\frac{\tau p^4}{2\gamma A}}
   \int_{\sqrt{\frac{\tau}{2\gamma A}} p^2}^{+\infty} dx\;
                            e^{-x^2}
   \stackrel{\Lambda\rightarrow 0}{\approx}                 \nonumber\\
&& \sqrt{ \frac{\pi\tau}{\gamma A} }
   \int_{\Lambda} p\;dp\;U(p)=
   -\sqrt{\frac{\pi\tau}{\gamma A}}\;\ln(\Lambda) \;\;.
\end{eqnarray}
The regularized asymptote of the density of vortices is

\begin{equation}
n_{REG.}\approx \frac{1}{2\pi^{3/2}} \sqrt{\frac{\gamma A}{\tau}}\;\;.
\end{equation}
The $1/2$ critical exponent is consistent with the numerical results 
in \cite{zy}.

\section{ Density of monopoles. }

  The analysis of monopoles in $D=3$ dimensions is much the same as that
of vortices in $D=2$ dimensions. The monopole configuration is

\begin{equation}
S^{\alpha}(\vec{x})=f(r) [ \delta^{\alpha 1}\sin(\theta)\cos(\varphi)+
                           \delta^{\alpha 2}\sin(\theta)\sin(\varphi)+
                           \delta^{\alpha 3}\cos(\theta)       ] \;\;,
\end{equation}
where $r,\theta,\varphi$ are spherical coordinates. The profile function
interpolates between $f(0)=0$ and $f(\infty)=1$. A single monopole
has infinite energy, the IR divergence is due to the long distance
asymptote

\begin{equation}\label{asymptmonopole}
S^{a}_{\infty}(\vec{x})=[ \delta^{a1}\sin(\theta)\cos(\varphi)+
                          \delta^{a2}\sin(\theta)\sin(\varphi)+
                          \delta^{a3}\cos(\theta)                 ] \;\;.
\end{equation}
The asymptote is scale invariant. The asymptote determines the
long distance behaviour of the Fourier transform, which is

\begin{equation}\label{Umonopole}
U(p)\sim \frac{1}{p^4} \;\;
\end{equation}
up to a numerical factor. Similarly as for vortices this singularity
alone determines the critical exponents.

\subsection{Underdamped limit.}

  Combination of the formulas (\ref{Gfrozenunder},\ref{n},\ref{Umonopole})
gives the density of monopoles for large $\tau$. The numerator
integral in (\ref{n}) is proportional to

\begin{eqnarray}
&& \int_{\Lambda}^{+\infty} p^2\;dp\;p^2\; U(p)\; G(\tau,p)
   \stackrel{\tau\rightarrow\infty}{\approx}                       \nonumber\\
&& \int_{\Lambda}^{+\infty} p^2\;dp\;p^2\; U(p)\;
   \frac{1}{\frac{2A}{\gamma\tau}+p^2}
   \stackrel{\tau\rightarrow\infty}{\approx}                       \nonumber\\
&& \int_{\Lambda}^{\infty} p^2\;dp\; U(p)\;
   \stackrel{\Lambda\rightarrow 0}{\approx} \frac{1}{\Lambda}      \;\;.
\end{eqnarray}
$\Lambda$ is an IR regulator which is provided either by the system
size or by an average distance between monopoles and antimonopoles. The
denominator integral is proportional to

\begin{eqnarray}
&& \int_{\Lambda}^{+\infty} p^2\;dp\; U(p)\; G(\tau,p)
   \stackrel{\tau\rightarrow\infty}{\approx}                  \nonumber\\
&& \int_{\Lambda}^{+\infty} p^2\;dp\; U(p)\;
   \frac{1}{\frac{2A}{\gamma\tau}+p^2}
   \stackrel{\Lambda\rightarrow 0}{\approx}                   \nonumber\\
&& \frac{\gamma\tau}{2A} \int_{\Lambda} \frac{dp}{p^2}=
   \frac{\gamma\tau}{2A} \frac{1}{\Lambda}                    \;\;.
\end{eqnarray}
We manipulated with the order of the limits $\tau\rightarrow\infty$
and $\Lambda\rightarrow 0$ so as to get rid of the IR divergence
and obtain a finite regularized asymptote

\begin{equation}
n_{REG.} \approx \frac{1}{\pi^2}
                 (\frac{2A}{\gamma\tau})^{\frac{3}{2}}  \;\;.
\end{equation}
The asymptote (\ref{asymptvortex}) is scale invariant; the density of 
monopoles scales like an inverse qubed of the frozen correlation 
length (\ref{xifrozenunder}).

\subsection{Overdamped limit.}

   In this limit we combine the formulas
(\ref{Gasymptover},\ref{n},\ref{Umonopole}). The numerator integral in
Eq.(\ref{n}) is proportional to

\begin{eqnarray}
&& \int_{\Lambda}^{+\infty} p^2\;dp\;p^2\; U(p)\; G(\tau,p)
   \stackrel{\tau\rightarrow\infty}{\approx}                \nonumber\\
&& \int_{\Lambda}^{+\infty} p^2\;dp\; U(p)\;
   \stackrel{\Lambda\rightarrow 0}{\approx}
   \;\frac{1}{\Lambda} \;\;.
\end{eqnarray}
The denominator is

\begin{eqnarray}
&& \int_{\Lambda}^{+\infty} p^2\;dp\; U(p)\; G(\tau,p)
   \stackrel{\tau\rightarrow\infty}{\approx}                \nonumber\\
&& \sqrt{\frac{2\tau}{\gamma A}}
   \int_{\Lambda}^{+\infty} p^2\;dp\;U(p)\;
                            e^{\frac{\tau p^4}{2\gamma A}}
   \int_{\sqrt{\frac{\tau}{2\gamma A}} p^2}^{+\infty} dx\;
                            e^{-x^2}
   \stackrel{\Lambda\rightarrow 0}{\approx}                 \nonumber\\
&& \sqrt{ \frac{\pi\tau}{\gamma A} }
   \int_{\Lambda} p^2\;dp\;U(p)=
   \sqrt{\frac{\pi\tau}{\gamma A}}\;\frac{1}{\Lambda} \;\;.
\end{eqnarray}
The regularized $\tau>>2\gamma A$ asymptote of the density of vortices is

\begin{equation}
n_{REG.}\approx \frac{1}{\pi^{11/4}}
                (\frac{\gamma A}{\tau})^{\frac{3}{4}} \;\;.
\end{equation}

\section{ Conclusion. }

   We considered density of hedgehogs just after a rapid pressure
quench. For large enough quench time $\tau$ the density depends
on $\tau$ like

\begin{equation}
n\sim \tau^{-\sigma} \;\;.
\end{equation}
Our findings are summarized in a table.

\vspace{0.5cm}
\begin{tabular}{||c|c|c|c||}
\hline\hline
    &          &          \multicolumn{2}{c|}{$\sigma$}            \\
\cline{3-4}
$D$ &  DEFECT  & $\gamma\rightarrow 0$ & $\gamma\rightarrow\infty$ \\
\hline
$2$ &  VORTEX  &         $1$           &     $\frac{1}{2}$         \\
\hline
$3$ & MONOPOLE &    $\frac{3}{2}$      &     $\frac{3}{4}$         \\
\hline\hline
\end{tabular}
\vspace{0.5cm}

   The critical exponent depends on the degree of damping. The underdamped
critical exponent is twice the overdamped critical exponent for a given
defect. It is not surprising as the two limits are dominated by a second
order and a first order time derivative respectively.

   Global vortices and global monopoles interact by long distance confining
potentials. Their density just after a quench is merely an initial condition
for their subsequent rapid recombination described by phase ordering
kinetics \cite{bray}. Their critical exponents apply to local
vortices/monopoles coupled to a gauge field provided that the gauge
field equilibrates with the matter field rather slowly \cite{zy}.
These local defects have finite energy and interact weakly so their
subsequent recombination is purely diffusive.

   Topological textures like Bloch waves for $D=1$, skyrmions for
$D=2$, magnetic bubbles and Hopfions for $D=3$ require a separate study
\cite{bloch}.

\end{document}